\documentstyle[amssymb,12pt,graphics,aps]{revtex}
\begin{document}\draft
\title{\bf Explicit Spectral formulas for scaling quantum graphs}
\author{Yu. Dabaghian$^1$ and R. Bl\"umel$^2$}
\address{$^1$Department of Physiology,
Keck Center for Integrative Neuroscience,\\
University of California,
San Francisco, California 94143-0444, USA}
\address{$^2$Department of Physics, Wesleyan University,
Middletown, Connecticut 06459-0155, USA}
\date{\today}
\maketitle
\begin{abstract}
We present an exact analytical solution of the
spectral problem of quasi one-dimensional scaling
quantum graphs. Strongly stochastic in the classical
limit, these systems are frequently employed
as models of quantum chaos.
We show that despite their classical stochasticity
all scaling quantum graphs are explicitly solvable
in the form $E_n=f(n)$, where $n$ is the sequence number
of the energy level of the quantum graph
and $f$ is a known function, which depends only
on the physical and geometrical properties of the
quantum graph.
Our method of solution
motivates a new classification
scheme for quantum graphs:
we show that each quantum graph can be uniquely assigned an
integer $m$ reflecting its level of complexity.
We show that a network of taut strings with piecewise constant
mass density
provides an experimentally realizable analogue system
of scaling quantum graphs.

\end{abstract}



\section{Introduction}
Quantum graphs\cite{QGT1,QGT2,QGreview} are the
``harmonic oscillators'' of quantum chaos.
Due to their structural simplicity
they provide a test bed
for a large number of properties
and hypotheses of
quantum chaotic systems. Many theoretical investigations,
which are difficult to conduct for
more familiar quantum chaotic systems \cite{Gutzw,Haake,STOECK},
can be carried out explicitly for quantum graphs,
both in the classical and in the quantum regimes.
An example are
recently obtained spectral formulas \cite{Opus,Prima,Sutra,Stanza},
which provide
explicit analytical expressions for the {\it individual}
quantum energy eigenvalues of a subset
of scaling quantum graphs.

Recently we were able to generalize our methods
to the set of {\it all} scaling quantum graphs
\cite{PREthatis}. The purpose of this
paper is to provide a more detailed
discussion
and to present new results on
the spectral statistics and
the convergence of
our explicit solution formulas.
We also present a new classification
scheme of scaling quantum graphs.
We show that it is possible to
label each scaling quantum graph
with
an integer $m$ which reflects the
degree of complexity
of its spectrum.
We also suggest an experimentally realizable
analogue system of scaling quantum graphs.
This shows that scaling quantum graphs are
more than academic constructs, and that
physical systems can be found which can
be analyzed on the basis of
the theory of scaling
quantum graphs.
This view is corroborated by a recently published
microwave realization of quantum graphs \cite{Grexp}.

Our paper is organized in the following way.
In Sec.~II we introduce scaling quantum graphs and review
briefly explicit spectral formulas obtained for a sub-class
of scaling quantum graphs. In Sec.~III we examine the spectral
equation of scaling quantum graphs. In Sec.~IV we define
spectral separators whose knowledge enables the construction
of explicit spectral formulas for scaling quantum graphs.
We also define a new spectral hierarchy of scaling quantum
graphs which is based on the complexity of their spectra.
In Sec.~V we investigate the spectral statistics of
quantum graphs. We show that because of the existence of a
spectral cut-off the spectral statistics of finite quantum
graphs are never exactly Wignerian. We investigate the spectral
statistics of
a four-vertex scaling quantum graph in detail.
Comparing its spectral statistics with the spectral statistics
of more highly connected quantum graphs we show
that the index $m$, although indicative of the complexity of
the spectrum of a quantum graph, does not uniquely
characterize its spectral statistics.
In Sec.~VI we present Lagrange's inversion formula as
a new and alternative method for obtaining explicit spectral formulas.
In Sec.~VII we discuss our results. In Sec.~VIII we
summarize our results and conclude the paper. The paper
has two appendices. In Appendix A we provide a simple
proof for the statement that the spectral equation of
the $m=0$ complexity sub-class of
scaling quantum graphs has one and only one root per
root cell. This is important since our theory of
explicit spectral formulas of scaling quantum graphs
crucially hinges on this statement.
In Appendix B we show that our spectral
formulas are indeed convergent, and in addition that
they converge to the correct spectral points.

\section{Scaling quantum graphs}
As illustrated in Fig.~1, quantum graphs consist of a
quantum particle moving on a one-dimensional
network of bonds and vertices.
\begin{figure}
\begin{center}
\includegraphics{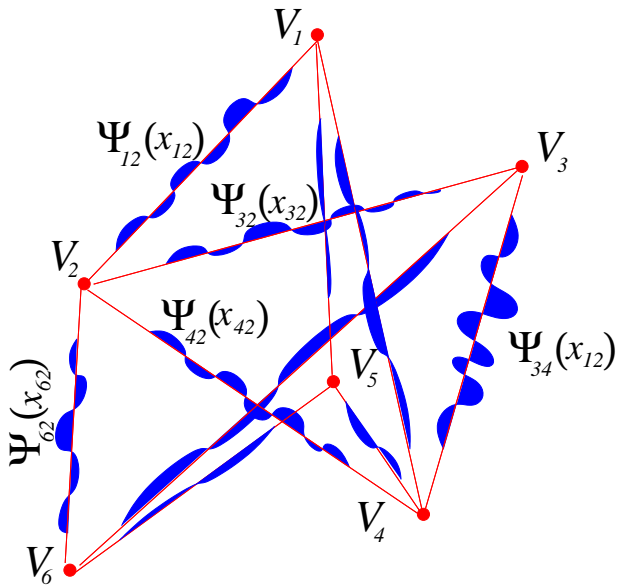}
\caption{\label{Fig.1} Quantum graph: a quantum particle moves
along the bonds of
a generic graph and scatters at its vertices.}
\end{center}
\end{figure}
The bonds $B_{ij}$ of the graph may be equipped with
potentials $U_{ij}$. We refer to these potentials as
bond potentials or the
{\it dressing} of the graph bonds. The parameters determining
the strength and the shape of the bond potentials are referred to
as {\it dressing parameters}.
In what follows the bond
potentials are considered to be {\it scaling} potentials,
$U_{ij}=\lambda_{ij}E$,
$\lambda_{ij}=\lambda_{ji}=const$. The physical meaning and the
reason for introducing the scaling assumption
are discussed in \cite{Opus,Prima,Sutra,Stanza}.
In addition, in Sec.~VII,
we present a physical analogue system
of scaling quantum graphs, a network of taut strings,
which has the same spectral equation as scaling quantum graphs.
The string system is an example of a naturally scaling system.
In a more general context
one can consider
the scaling assumption
as a tool which allows to avoid unnecessary mathematical
complications. For most physical systems scaling can be
achieved, even experimentally \cite{Tom}, by an appropriate choice of
parameters.
We also define $E=k^2$
since for the discussion below it is frequently
more convenient to
work with $k$ than to work with $E$.

For $\hbar=0$ quantum graphs produce strongly stochastic (mixing)
classical counterparts -- a
{\it classical} particle moving on the same one-dimensional network,
scattering randomly on its
vertices \cite{QGT1,QGT2,Orsay,Nova,Gaspard}.
We use the word
{\it stochastic} to characterize the
classical dynamics of the particle on the graph
since classically the scattering at the vertices is not
a deterministic process as required for
deterministic chaos\cite{Schuster}, but a random, stochastic
process, where the classical scattering probabilities are
determined directly from the
quantum dynamics in the limit $\hbar\rightarrow 0$
\cite{RS2}.

Despite the apparent simplicity of quantum graphs,
their behavior exhibits many
familiar features of classically chaotic systems.
Examples are the exponential
proliferation of classical periodic orbits and the
approximate Wignerian statistics
of nearest-neighbor spacings \cite{QGT1,QGT2} (see also Sec.~V).
As a result quantum graphs are {\it quantum stochastic}
systems, which mimic closely
the behavior of quantum chaotic systems. It is therefore
very interesting that despite
their classical stochasticity and despite many familiar
phenomenological features of
quantum chaos exhibited in the quantum regime, the spectral
problem for scaling quantum
graphs turns out to be explicitly solvable \cite{PREthatis,Latest}.

Let us first outline the solution for a particular
class of scaling quantum graphs,
called {\it regular} in \cite{Opus,Prima,Sutra,Stanza}.
We note that the term ``regular''
as used here refers to the regular behavior of the spectrum
of the corresponding quantum
graphs and has nothing to do with regular graphs
as defined in graph theory \cite{GrTheo},
e.g. graphs with a fixed coordination number.
A case in point is the recent paper by
Severini and Tanner \cite{SevTan} where the term
``regular quantum graphs'' refers to quantum graphs with
a special graph topology.

For regular quantum graphs
there exists a set of $k$-intervals $I_{n}$, each of
which contains precisely
one momentum eigenvalue $k_n$ (see Appendix~A).
The end points $\hat k_{n}$ of these intervals,
$I_{n}=\left[\hat k_{n-1}, \hat k_{n}\right]$, form a periodic set,
\begin{equation}
\hat k_{n}=\kappa_1 n+\kappa_2,
\label{kappa}
\end{equation}
where the constants
$\kappa_1,\kappa_2$ are
determined explicitly
in terms of the parameters
of the quantum graph. Clearly, the points $\hat k_{n}$
separate the eigenvalues
$k_{n}$ from each other, and are therefore called {\it separators}
(see Sec.~IV).

As soon as the separators $\hat k_n$ and the density of states
$\rho(k)$
are known,
an explicit expression for the
energy eigenvalues of a given quantum graph is obtained
either by first computing
the momentum eigenvalues
\begin{equation}
k_{n}=\int_{\hat k_{n-1}}^{\hat k_{n}}\rho(k)\,kdk ,
\label{regularroots}
\end{equation}
and then using $E_n=k_n^2$, or by computing $E_{n}$ directly as
\begin{equation}
E_{n}=\int_{\hat E_{n-1}}^{\hat E_{n}}\rho(E)\,EdE ,
\label{regularE}
\end{equation}
where $\hat E_{n}=\hat k_{n}^{2}$, $\rho(E)dE=\rho(k)dk$.
An explicit periodic-orbit
expansion of the density of states $\rho(k)$ is given by \cite{Opus,Prima}
\begin{equation}
\rho(k)\equiv\sum_{n}\delta(k-k_{n})=
\frac{S_{0}}{\pi}+\mathop{\rm Re}\frac{1}{\pi}
\sum_{p} S_{p}^0
\sum_{\nu =1}^{\infty}A_{p}^{\nu}e^{i\nu S^{0}_{p}k},
\label{rho}
\end{equation}
where $S^{0}_{p}$, and $A_{p}$ are correspondingly
the reduced action lengths
and the weight factors of the prime periodic
orbits labeled by $p$, $\nu$ is the
multiple traversal index, and $S_{0}$ is
the total reduced action length of the
graph \cite{Sutra}.
The constant term in the expansion (\ref{rho})
of $\rho$ shows that $\kappa_1$ in Eq.~(\ref{kappa})
is given by $\kappa_1=\pi/S_0$. In order to
illustrate the construction of explicit spectral
formulas we assume, for simplicity, that
$\kappa_2=1/2$ and all $A_p$ are real.
Both assumptions hold
for a large class of regular quantum graphs.
If we now use
the expansion (\ref{rho}) in Eq.~(\ref{regularE})
we arrive at the following exact,
explicit periodic-orbit expansion of the individual
energy levels of the corresponding regular quantum graphs:
\begin{eqnarray}
E_{n} &=&\frac{\pi^{2}}{S_{0}^{2}}\left( n^{2}+\frac{1}{12}\right)
-\frac{%
4\pi n}{S_{0}^{2}}\mathop{\rm Im}\sum_{p,\nu }
\frac{A_{p}^{\nu }}{%
\omega_{p}\nu ^{2}}\sin \left(\frac{\omega _{p}\nu }{2}\right)\,
e^{in\omega _{p}\nu }\cr &&-%
\frac{4\pi }{S_{0}^{2}}\mathop{\rm Re}\sum_{p,\nu }
\frac{A_{p}^{\nu }}{\nu
^{3}\omega _{p}^{2}}\left[ \sin \left(\frac{\omega _{p}
\nu }{2}\right)-\left(\frac{\omega
_{p}\nu }{2}\right)
\cos \left(\frac{\omega _{p}\nu }{2}\right)\right]\,
e^{in\omega _{p}\nu },
\label{regroots}
\end{eqnarray}
where $\omega_{p}=\pi S_{p}^{0}/S_{0}$.
Therefore, according to (\ref{regroots}), the
index $n$ that counts the separators
$\hat k_{n}$ of the regular
quantum graph, is a {\it quantum number}
in the sense that it explicitly enumerates
the physical eigenstates.
In this respect, the explicit formulas for
the quantum energy levels $E_{n}$ of these
systems are analogous to the well-known
Einstein-Brillouin-Keller (EBK) quantization
formulas for integrable systems \cite{Gutzw,Haake,STOECK}.
This is a very interesting fact from
the point of view of the semiclassical periodic-orbit
quantization theory. In this
respect, the regular quantum graphs represent curious
hybrids of classical stochasticity
and quantum spectral solvability.

However, the systems for which the expansion (\ref{regroots})
is valid, represent a very
special class of quantum graphs. Just how
special such ``spectral regularity'' is, can be
illustrated in terms of the behavior of the
corresponding spectral staircase function,
\begin{equation}
N(E)=\sum_{n}\theta(E-E_{n}),
\label{staircase}
\end{equation}
where $\theta$ is the unit
step function defined as
\begin{equation}
\theta(x)=\cases{0, &for $x<0$, \cr
                1/2, &for $x=0$, \cr
                 1, &for $x>0$.\cr}
\label{Heavy}
\end{equation}
It was shown in \cite{Opus,Stanza},
that for the regular systems, the average spectral staircase
(Weyl's average),
\begin{equation}
\bar N(E)=\frac{S_{0}}{\pi}\sqrt{E}+\bar N (0),
\label{Nave}
\end{equation}
has the {\it piercing property}, i.e. it
intersects every stair step of the spectral
staircase function $N(E)$, as illustrated in Fig.~2.
\begin{figure}
\begin{center}
\includegraphics{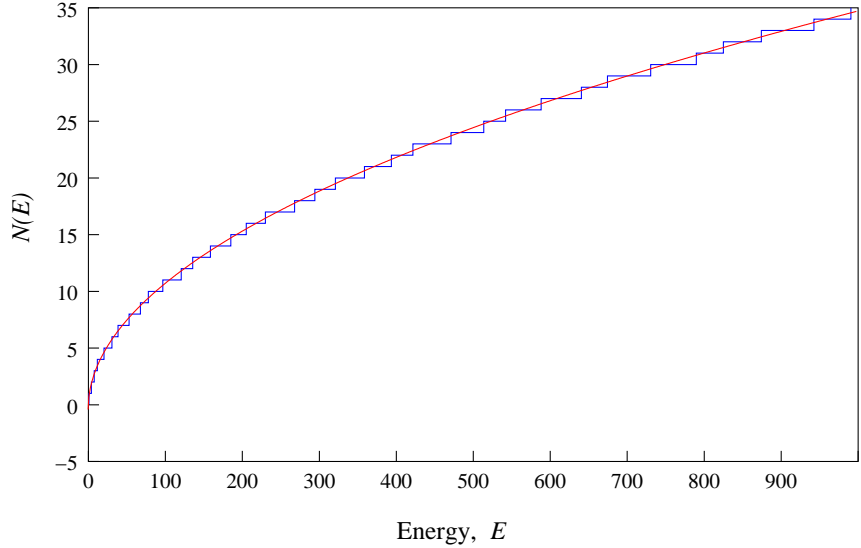}
\caption{\label{Fig.2} Piercing property of the regular quantum graphs.
The spectral staircase function of a regular
quantum graph is pierced by its average $\bar N(E)$.}
\end{center}
\end{figure}
If a quantum system has the piercing property,
there exists exactly one intersection
point $\hat E_{n-1}$, between every two neighboring energy
levels $E_{n-1}< E_{n}$,
\begin{equation}
N(\hat E_{n-1})=\bar N (\hat E_{n-1}),\ \ E_{n-1}< \hat E_{n-1}< E_{n},
\ \ \ n=1,2,\ldots\,\, .
\label{pierce}
\end{equation}
The $\hat E_n$ thus defined may serve as separators for
the quantum energy spectrum.
As shown in Fig.~2 the
piercing-average requirement (\ref{pierce})
is indeed quite restrictive.
Consequently, regular quantum graphs form a relatively small subset
of quantum graphs.
As demonstrated in \cite{Stanza,Wilmington}, only a few
graph topologies (for instance
linear chains) admit a regular regime for an appropriate
choice of network parameters.
As an example, a four-vertex linear-chain quantum graph
(see inset of Fig.~3),
which is characterized by the values
of the two reflection coefficients $r_{2}$ and $r_{3}$ at
the two middle vertices $V_{2}$ and
$V_{3}$, is in the regular regime if these
parameters fall into the shaded region shown in
Fig. 3.
\begin{figure}
\begin{center}
\includegraphics{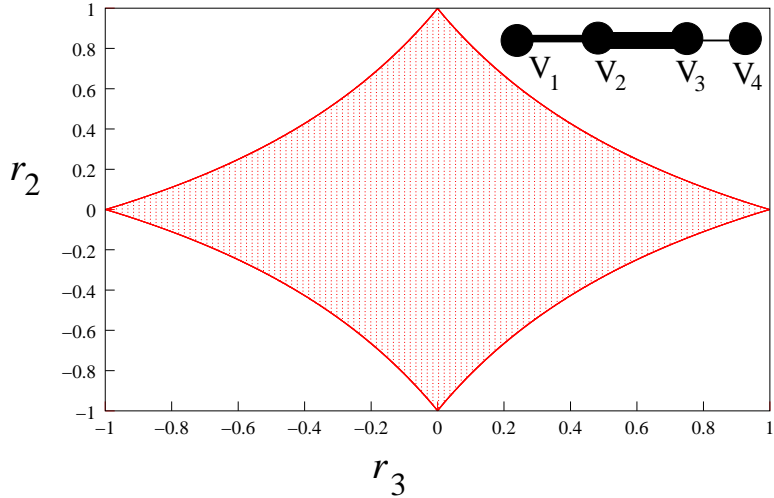}
\caption{\label{Fig.3} The parameter space of the four-vertex
linear graph. The
shaded region corresponds to the regular regime.}
\end{center}
\end{figure}
The majority of scaling quantum graphs do not admit
regular regimes. Hence it is intriguing
to understand the spectral behavior of irregular
quantum graphs, i.e.
those for which the piercing-average
condition (\ref{pierce}) is violated.

\section{Spectral equation}
In order to set the stage for the following discussion,
let us recall some general
definitions and properties of quantum graphs.
As mentioned in the introduction, a quantum
graph \cite{QGT1,QGT2,QGreview} consists of a
quantum particle moving on
a one-dimensional network of $N_{B}$ bonds connecting $N_{V}$
vertices (Fig.~1). Every bond $B_{ij}$ which
connects the vertices $V_{i}$ and $V_{j}$, carries a
solution of the Schr\"{o}dinger equation,
$\hat H\psi_{n}=E_{n}\psi_{n}$. The length of the
bonds is denoted by $L_{ij}$.
With the
constant {\it scaled} potentials $U_{ij}=\lambda_{ij}E$
defined on the bonds of the graph,
the Schr\"odinger equation is
\begin{equation}
\frac{d^{2}}{dx^{2}_{ij}}\psi_{ij}(x)+\beta_{ij}^{2}E\psi_{ij}(x)=0,
\label{schred}
\end{equation}
where $\beta_{ij}=\pm\sqrt{1-\lambda_{ij}}$.

Below we shall assume for simplicity that the
energy $E$ is kept above the maximal scaled potential
height, i.e. $\lambda_{ij}<1$, $i,j=1,...,N_{V}$,
so that tunneling solutions
are excluded and the general solution of
Eq.~(\ref{schred}) on the bond $B_{ij}$ is
\begin{equation}
\psi_{ij}(x)=a_{ij}e^{-ik\beta_{ij}x_{ij}}+b_{ij}e^{ik\beta_{ij}x_{ij}}.
\label{psi}
\end{equation}
The quantization conditions for quantum
graphs are the result of the requirement that the solutions
(\ref{psi}) must satisfy the continuity
and the current conservation conditions at every vertex
$V_{i}$. The procedure of imposing the boundary
conditions can be reformulated in terms of an
auxiliary problem of quantum scattering on the
vertices of the graph \cite{QGT2,Stanza,Orsay},
which provides an elegant solution of the graph quantization
problem.
As shown in \cite{QGT2,Stanza,Orsay} the consistency of the
complete set of boundary conditions
at all vertices yields the spectral equation
\begin{equation}
\Delta (k)=\det\left[1-S(k)\right] =0,
\label{det}
\end{equation}
where $S(k)$ is a $2N_{B}\times 2N_{B}$
unitary (scattering) matrix \cite{QGT2,Stanza,Orsay},
\begin{equation}
S_{IJ}(k)=t_{IJ}e^{i\beta_{I}L_{I}k}.
\label{S}
\end{equation}
Here the capital indices $I$, $J$ are used to denote the
directed bonds, $I,J=1,2,...,2N_{B}$. We denote by $I'$ the
time-reversed bond of $I$.
The elements $t_{IJ}$ (discussed in detail in \cite{Stanza})
have the meaning of transmission (reflection)
amplitudes for transitions between the
(directed) bonds $I$ and $J$. Transmission occurs if
$I$ and $J$ are connected and $J\neq I'$.
If $I$ and $J$ are not connected, we have $t_{IJ}=0$.
An example here is $t_{II}=0$ for all $I$.
For $J=I'$ the matrix element $t_{IJ}$ has the meaning of a
reflection amplitude
\cite{QGT2,Sutra,Stanza,Orsay}.
Due to the scaling condition, the $t_{IJ}$'s are
constant ($k$-independent) parameters.

For conventional quantum graphs without potential dressing
the connection between
the coefficients $t_{IJ}$
and the expansion coefficients $A_{p}$ in
Eqs.~(\ref{rho}) and (\ref{regroots}) was established
early on in the seminal literature on quantum graphs, e.g. in
Refs.~\cite{QGT1,QGT2}.
Later it was shown to hold also in the case of dressed,
scaling quantum graphs \cite{Nova}.
Each transition of an orbit $p$ from a
bond $I$ to $J$ contributes the
factor $t_{IJ}$ to the weight
$A_{p}$ of the orbit, so that
\begin{equation}
A_{p}=\prod_{\{p\}}t_{IJ},
\end{equation}
where the product is taken over the
sequence of bonds traced.

Note that the phases of the exponentials in Eq.~(\ref{S})
coincide with the classical actions associated
with the particle path traversing the bond $B_{I}$,
\begin{equation}
S_{I}(k)=\beta_{I}L_{I}k.
\label{action}
\end{equation}
The spectral determinant (\ref{det}) can be
written in the form
\begin{equation}
\Delta(k) = e^{i\Theta_0(k)}\, \Delta_R(k),
\label{ReDelta}
\end{equation}
where $\Delta_R(k)$ is the (real) modulus of
$\Delta(k)$ and $\Theta_0(k)$ is its phase.
The phase is given by
\cite{Sutra}
\begin{equation}
\Theta_{0} (k)=
\frac{1}{2}\ln\det S=S_{0}k-\pi\gamma_{0},
\label{theta}
\end{equation}
where $S_0$,
the total reduced action length as introduced in Eq.~(\ref{rho}),
is given explicitly by
\begin{equation}
S_{0}={1\over 2}\sum_{I=1}^{2N_{B}}L_{I}\beta_{I}
\label{length}
\end{equation}
and $\gamma_0$ is a constant phase.
The modulus is given by \cite{Sutra}
\begin{equation}
\Delta_R(k)=
\cos\left(S_{0}k-\pi\gamma_{0}\right)-
\sum_{i=1}^{N_{\Gamma}}a_{i}\cos (S_{i}k-\pi\gamma_{i}),
\label{cos1}
\end{equation}
where $a_{i}$ are constant coefficients, $\gamma_{i}$
are constant phases, $N_{\Gamma}$ is the number
of harmonic terms in the sum of Eq.~(\ref{cos1})
and the frequencies $S_{i}$ are linear
combinations of the reduced classical bond
action lengths $S_{I}^{0}=\beta_{I}L_{I}$.
$S_0$
is the largest frequency in Eq.~(\ref{cos1}),
i.e. $S_i<S_0$, $i=1\ldots N_{\Gamma}$. This
fact will be of crucial importance below.

The spectrum of the quantum graph is obtained from the equation
\begin{equation}
\Delta_R(k)=0.
\label{cos}
\end{equation}
In Appendix A we prove that
if the coefficients of the {\it characteristic
function}  $\Phi(k)$ of the graph,
\begin{equation}
\Phi(k)\equiv \sum_{i=1}^{N_{\Gamma}}a_{i}\cos (S_{i}k-\pi\gamma_{i}),
\label{char}
\end{equation}
satisfy the condition
\begin{equation}
\sum_{i=1}^{N_{\Gamma}} \left| a_{i}\right| \equiv \alpha <1,
\label{reg}
\end{equation}
precisely one
solution $k_{n}$ of Eq.~(\ref{cos}) can be found between
each two sequential separators
\begin{equation}
\hat{k}_{n}=\frac{\pi}{S_{0}}\left(n+\gamma_{0}+\mu+1\right),
\label{sep}
\end{equation}
where $\mu$, an integer, is to be adjusted such that
$k_1<\hat k_1<k_2$.
This is the case, e.g. for a two-bond graph
(Fig.~4) with the bond lengths $L_{1}$ and $L_{2}$,
for which the spectral equation is
\begin{equation}
\sin (S_{0}k)-r\sin (S_{1}k)=0.
\label{3hydra}
\end{equation}
Here $S_{0}=L_{1}\beta_{1}+L_{2}\beta_{2}$, $S_{1}=
L_{1}\beta_{1}-L_{2}\beta_{2}$,
and $r$ is a constant positive reflection
coefficient at the vertex $V_{2}$ between
the two bonds. Since $|r|< 1$, the condition (\ref{reg}) is satisfied
and hence this graph is always regular.
\begin{figure}
\begin{center}
\includegraphics{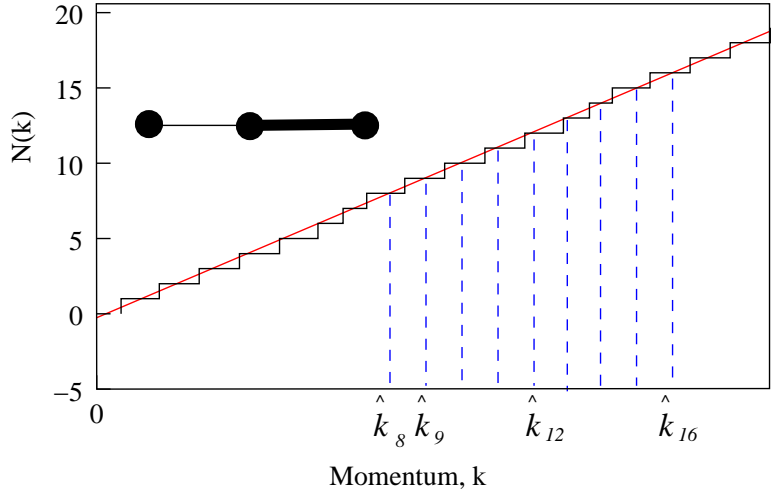}
\caption{\label{Fig.4} The three-vertex linear graph
(inset) and
the corresponding
staircase function. The intersections between $\bar N(k)$ and $N(k)$
correspond to the separating points $\hat k_{n}$.}
\end{center}
\end{figure}
In this case every step of the spectral
staircase function (\ref{staircase})
is pierced by its average (Fig.~4), or equivalently,
every interval
$I_{n}=[\hat{k}_{n-1},\hat{k}_{n}]$ contains precisely one
quantum eigenvalue of the momentum.
This spectral regularity is the key for
obtaining the explicit harmonic expansion for
each individual root of the spectral
determinant (\ref{det}).
In general, however, the regularity
condition (\ref{reg}) does not hold and hence
the principle ``one root per interval $I_{n}$''
(see Appendix~A) is
violated. This is illustrated in
Fig.~5, which shows the behavior of the spectral
staircase for the four-vertex
linear chain in two different dynamical regimes.
The spectral staircase on the right corresponds
to a case in which the parameters $r_{2}$ and $r_{3}$
fall outside of the shaded regularity
region in Fig.~3.
\begin{figure}
\begin{center}
\includegraphics{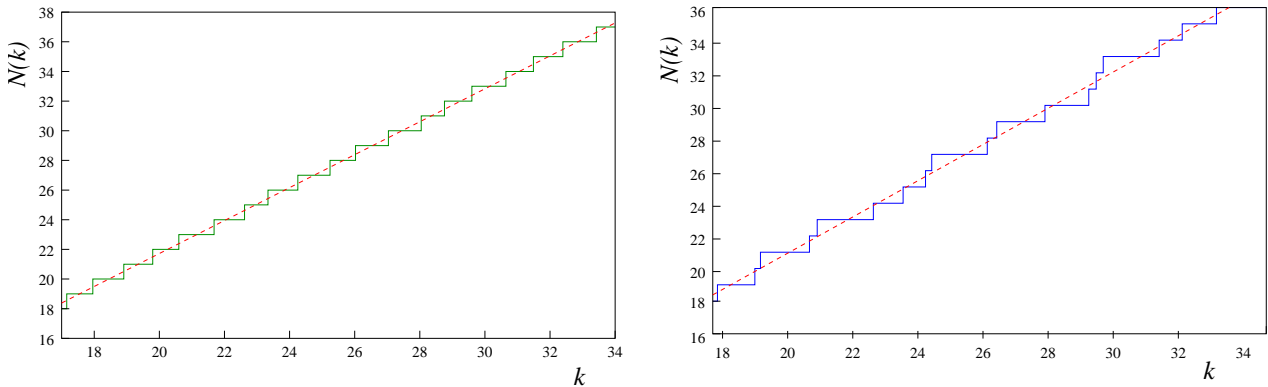}
\caption{\label{Fig.5} The staircase and the
average $\bar N(k)$ for the four-vertex
linear graph in a regular (left) and in an irregular (right) regime.}
\end{center}
\end{figure}
Hence, in order to proceed with an analysis
similar to the one for regular quantum graphs,
one needs to find a set of separating points
that ``bootstrap'' the spectrum, and allow us
to integrate around each
delta-peak of $\rho(k)$, as in Eq.~(\ref{regularroots}).

\section{Separators}
What is the set of points that can be used as
separators for a generic quantum graph? Since the points
$k_{n}$ that need to be separated, are the zeros of
the spectral determinant $\Delta_R(k)$,
one can invoke an elementary,
classic theorem \cite{Rolle}, which states
that between every two roots of
a real, continuous, differentiable
function there exists an extremum point.
Moreover, extending $\Delta_R(k)$ into the complex plane,
$\Delta_R(k)\rightarrow \Delta(z)$, and
using the Hadamard representation
of the resulting entire function $\Delta(z)$,
\begin{equation}
\Delta(z)=e^{i(S_{0}z-\pi\gamma_{0})}z^{q}
\prod_{n}\left(1-\frac{z}{k_{n}}\right)e^{\frac{z}{k_{n}}},
\label{hadamard}
\end{equation}
where $q\geq 0$ is the multiplicity of the
root $k_{n}=0$, and all the roots $k_n$ are assumed to be real
as required, since $\Delta_R(k)$ is derived from a Hermitian
eigenvalue problem,
one can show \cite{Laguerre,Levin} that there
is exactly one zero between every two neighboring
extrema of $\Delta_R(k)$, i.e. that the zeros and the
extrema {\it interlace} and ``extra wiggles''
such as, e.g., illustrated by the dashed line in
Fig.~6, are {\it not}
possible. Hence the locations of extrema may be used as the
separating points for bootstrapping the physical spectrum.
\begin{figure}
\begin{center}
\includegraphics{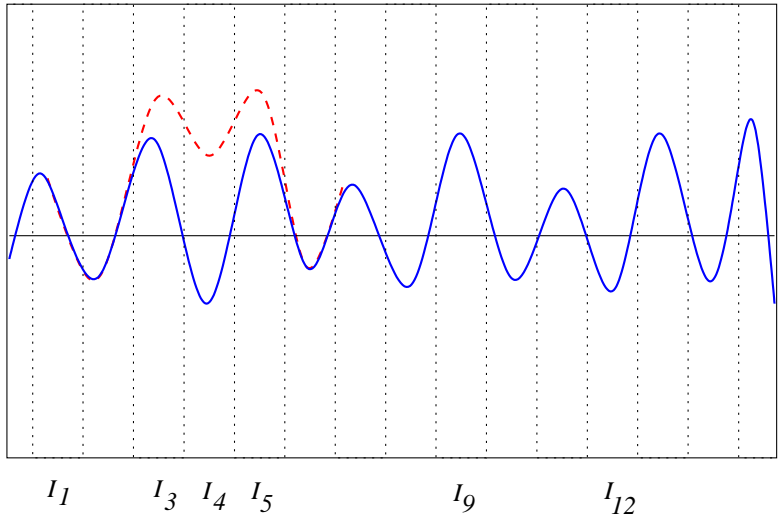}
\caption{\label{Fig.6} The interlacing
sequence of roots and extrema of the
spectral determinant. The dashed line represents
the forbidden ``extra wiggles''.}
\end{center}
\end{figure}
Strictly speaking, all this works only
for simple roots of $\Delta(z)$,
which is the generic case.
Multiple roots may, and in special cases do, occur.
But as explained in Refs.~\cite{PREthatis,Latest}, these
cases are trivial to deal with. In such a case the
separators and the spectral points are degenerate
and no further computation is necessary.

Is it any easier to obtain the extrema
of $\Delta_R(k)$ than to obtain its roots?
Interestingly, looking for
the answer to this question provides us with a
complete scheme for establishing a
hierarchy of quantum graphs according to
their spectral irregularity. Let us examine
this question more closely.

The equation $\Delta_R'(k)=0$ that defines the
extrema of $\Delta_R(k)$ is
\begin{equation}
\sin\left(S_{0}k-\pi\gamma_{0}\right)-
\sum_{i=1}^{N_{\Gamma}}a_{i}\epsilon_{i}\sin (S_{i}k-\pi\gamma_{i})=0,
\label{spectralder}
\end{equation}
where
\begin{equation}
\epsilon_{i}=\frac{S_{i}}{S_{0}}<1.
\label{factors}
\end{equation}
Note that this is the same type of
trigonometric polynomial as the original spectral equation
(\ref{cos}) (with a shifted phase $\gamma_{0}$),
with the new characteristic function
\begin{equation}
\Phi^{(1)}(k)\equiv\sum_{i=1}^{N_{\Gamma}}a_{i}
\epsilon_{i}\sin (S_{i}k-\pi\gamma_{i}).
\label{char1}
\end{equation}
However, compared to the original $\Phi(k)$ of Eq.~(\ref{char}),
this function has certainly a better chance
of satisfying the regularity condition (\ref{reg}), because
the absolute values of
the coefficients $a_{i}$ have been decreased by
the factors $\epsilon_{i}<1$, i.e.
\begin{equation}
a_{i}\rightarrow a_{i}\epsilon_{i}.
\label{coefficients}
\end{equation}
Let us assume that for a certain irregular graph with
$\sum_{i=1}^{N_{\Gamma}}\left| a_{i}\right|>1$, the new
characteristic function $\Phi^{(1)}(k)$
actually does satisfy the regularity condition (\ref{reg}), i.e.
\begin{equation}
\sum_{i=1}^{N_{\Gamma}}\left|a_{i}\epsilon_{i}\right| <1.
\label{reg1}
\end{equation}
According to the results of Secs.~II and III this means
that the zeros of
$\Delta_R'(k)$ (the extremal points of
$\Delta_R(k)$) can be separated from each other by the
periodic sequence of separators (\ref{sep}), i.e.
that there exists exactly one extremum of
$\Delta_R(k)$ between every two points
$\hat k_{n}$. For functions of the type (\ref{hadamard})
the converse statement is also true \cite{Laguerre,Levin}, i.e.
there exists a root of $\Delta_R(k)$
between every two extrema of $\Delta_R(k)$.
This suggests a direct strategy for
obtaining the roots of Eq.~(\ref{cos}).
First, as mentioned above,
we note that
the spectral equation for $\Delta_R'(k)$ can be written
in the form
\begin{equation}
\Delta_R'(k)=\sum_{j=0}^{N_{\Gamma}}C_j
\cos(S_jk+\varphi_j)=0,
\label{seD_R}
\end{equation}
where $C_j$ and $\varphi_j$ are constants. From
this we obtain the following explicit
formula for the {\it density-of-extremas}
functional $\rho^{(1)}(k)$
$$ \rho^{(1)}(k)=|\Delta_R''(k)|\, \delta(\Delta_R'(k))=
|\Delta_R''(k)|\, {1\over 2\pi}\, \int_{-\infty}^{\infty}\,
e^{i y \Delta_R'(k)}\, dy =$$
\begin{equation}
|\Delta_R''(k)|\, {1\over 2\pi} \sum_{n_0=-\infty}^{\infty}\ldots
\sum_{n_{N_\Gamma}=-\infty}^{\infty}
\left[\int_{-\infty}^{\infty}\prod_{j=0}^{N_{\Gamma}}
J_{n_j}(yC_j)\, dy\right]\,
\exp\left\{i\sum_{l=0}^{N_{\Gamma}}n_l[S_lk+\varphi_l]
\right\},
\label{rho1}
\end{equation}
where the $J_n$ are Bessel functions of
the first kind \cite{GR},
and the integrals in Eq.~(\ref{rho1}) converge for $C_j\neq 0$.
Now, using the expansion (\ref{rho1})
together with the periodic separators
(\ref{sep}), one obtains the separating
points $\hat k_{n}^{(0)}$ for the roots $k_{n}$ of
$\Delta_R(k)$, via
\begin{equation}
\hat k^{(0)}_{n}=\int_{\hat k_{n-1}^{(1)}}^{\hat k_{n}^{(1)}}
\rho^{(1)}(k)\,k\,dk.
\label{irregroots}
\end{equation}
Here we used the notation $\hat k_{n}^{(1)}$ for
the periodic separators (\ref{sep}), for future
convenience. Following this step, using the
separators $\hat k_{n}^{(0)}$ obtained in
Eq.~(\ref{irregroots}),
we find
the roots $k_{n}$ of the spectral equation via
\begin{equation}
k_{n}=\int_{\hat k_{n-1}^{(0)}}^{\hat k_{n}^{(0)}}
\rho^{(0)}(k)\,k\,dk,
\label{irregroots1}
\end{equation}
where the notation $\rho^{(0)}(k)$ was
used for the density of states $\rho(k)$.

For the case of the four-vertex linear graph, this
situation is again illustrated in Fig.~3, in which
it is now assumed that the differentiated
equation (\ref{spectralder}) satisfies the regularity condition
all through the domain $-1\leq r_{2},r_{3}\leq 1$.
This would be the case, e.g., if the bond
action lengths are chosen to be $S_{1}^0=0.25$, $S_{2}^0=0.45$,
and $S_{3}^0=1-S_{1}^0-S_{2}^0$.
For this case one can immediately
verify that the spectral equation
of the four-vertex linear chain,
\begin{eqnarray}
\sin(S_0 k)=r_3 \sin(S_1^0 k+S_2^0 k-S_3^0 k)-r_2 r_3
\sin(S_1^0 k-S_2^0 k+S_3^0 k)
\cr+
r_2\sin(S_1^0 k-S_2^0 k-S_3^0 k),
\label{4vertex}
\end{eqnarray}
is irregular outside of the region $|r_3|+|r_2r_3|+|r_2|<1$,
but the coefficients of the differentiated
equation satisfy the regularity condition (\ref{reg}).

Clearly this strategy can be
applied in the general case. If the differentiated equation
(\ref{spectralder}) is not of the regular type,
one can differentiate the spectral equation
(\ref{cos}) as many times as it is necessary
to obtain an equation of {\it regular type}
at the $m$-th step.
Indeed, the $l$-th derivative of the spectral equation is
\begin{equation}
\Delta_R^{(l)}(k)=
\cos\left(S_{0}k-\pi\gamma_{0}+\frac{\pi l}{2}\right)-
\sum_{i=1}^{N_{\Gamma}}a_{i}\epsilon_{i}^{l}
\cos\left(S_{i}k-\pi\gamma_{i}+\frac{\pi l}{2}\right).
\label{spectraldern}
\end{equation}
Obviously, since all $\epsilon_{i}$'s are
smaller than 1, we eventually (after a finite number $m$ of
steps) arrive at
an equation that satisfies the
regularity condition (\ref{reg}),
\begin{equation}
\sum_{i=1}^{N_{\Gamma}}\left|a_{i}\epsilon_{i}^{m}\right| <1.
\label{regreg}
\end{equation}
An upper bound for $m$ is easily established. We have
$m\leq -\ln(\sum_{i=1}^{N_{\Gamma}}|a_i|)/\ln(\max_i\epsilon_i)$.
Then, once the condition (\ref{regreg}) for
the $m$-th derivative of its spectral determinant is
satisfied, its zeros are separated from
each other by a periodic sequence of points,
\begin{equation}
\hat{k}_{n}^{(m)}=\frac{\pi}{S_{0}}\left(n+\gamma_{0}+\mu+1\right)
\label{sepr}
\end{equation}
as in Eq.~(\ref{sep}).
Using the density $\rho^{(m)}(k)$ of zeros of
$\Delta_R^{(m)}(k)$, which is obtained explicitly in complete
analogy with Eq.~(\ref{rho1}),
we can evaluate the zeros
themselves as
\begin{equation}
\hat k^{(m-1)}_{n}=\int_{\hat k_{n-1}^{(m)}}^{\hat k_{n}^{(m)}}
\rho^{(m)}(k)\,k\,dk.
\label{irregrootsr}
\end{equation}
Obviously, these points are now the
extrema of $\Delta_R^{(m-1)}(k)$, and moreover,
since there
is exactly one root of $\Delta_R^{(m)}(k)$ between
any two neighboring points $\hat{k}_{n}^{(m)}$,
$\Delta_R^{(m-1)}(k)$ has no
extrema other than the ones obtained in
Eq.~(\ref{irregrootsr}).

The newly obtained extrema of $\Delta_R^{(m-1)}(k)$
separate its zeros from each other, and
hence serve as the separators for the roots
of $\Delta_R^{(m-1)}(k)$. As a consequence, we can now
find {\it all} the roots of $\Delta_R^{(m-1)}(k)$
by using Eq.~(\ref{irregrootsr}) recursively
until we arrive at the $0$-th level
to obtain the roots $k_{n}$ of the
original spectral determinant.
This solves the problem of
obtaining the energy spectrum of any scaling quantum graph.

It is important to realize that a quantum graph of a 
certain topology can still have different degrees of irregularity
depending on the network's bond lengths and dressing parameters (for instance
the values of the bond potentials).
This point is easily illustrated by once more using the
example of the four-vertex linear chain.
Although the regularity region
for this graph is always the same,
the surrounding blank region in Fig.~3,
which corresponds to the irregular
regime, now acquires structure.
Figure~7 shows the spectral diagram
for the four-vertex linear-chain graph for two
different sets of graph parameters, corresponding
to two different irregularity regimes.
\begin{figure}
\begin{center}
\includegraphics{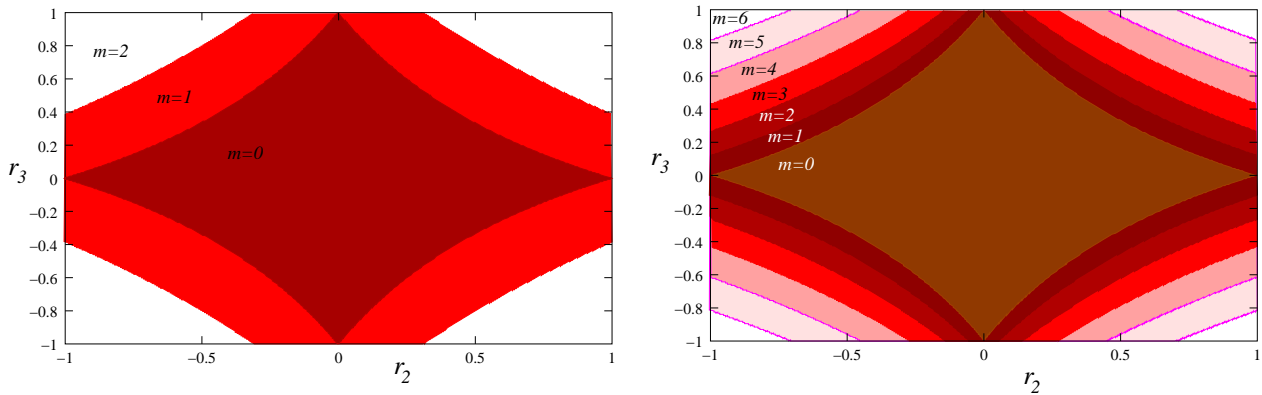}
\caption{\label{Fig.7}
The spectral regime diagram of the four-vertex linear-chain quantum graph.
The bond-action lengths are chosen to be $S_{1}^0=0.2$, $S_{2}^0=0.6565$,
$S_{3}^0=1-S_{1}^0-S_{2}^0$ (left panel), and $S_{1}^0=0.1$, $S_{2}^0=0.8565$,
$S_{3}^0=1-S_{1}^0-S_{2}^0$ (right panel) resulting in a maximum degree of
irregularity of $m=2$ (left panel) and $m=6$ (right panel).}
\end{center}
\end{figure}
The central diamond-shaped regions ($m=0$) in Fig.~7 are
the same as in Fig.~3 and correspond to the same regularity region as
in Fig.~3. The outer layers of the regular region in
Fig.~7
correspond to parameter values that
guarantee first ($m=1$), second
($m=2$), \ldots, degree of irregularity.

\section{Spectral statistics}
It is well known that the statistical
properties of the spectra of generic
quantum graphs are
well described by Random Matrix Theory (RMT)
\cite{QGT1,QGT2}. In particular,
the numerically obtained nearest-neighbor
distribution $P(s)$ of the
normalized spacings $s$ \cite{Haake} of the eigenvalues of
highly connected quantum graphs follows
closely the profile of the
Gaussian random matrix
ensembles, both in the presence of the time reversal
symmetry (GOE), where the nearest-neighbor distribution
is given by \cite{Haake}
\begin{equation}
P_{\rm GOE}(s)={\pi\over 2}\, s\, \exp(-s^2 \pi / 4)
\label{GOENNS}
\end{equation}
and in the
absence of it (GUE), where the nearest-neighbor distribution
is given by
\begin{equation}
P_{\rm GUE}(s)={32\over \pi^2}\, s^2\, \exp(-4s^2/ \pi).
\label{GUENNS}
\end{equation}
This circumstance is one of the most important
motivations for
studying quantum graphs in the context of quantum chaos theory.
In particular it is hoped to gain more insight into
the connection between chaos and random matrix theory
and, if possible, to prove the Bohigas-Giannoni-Schmit
conjecture \cite{BGS1,BGS2},
which states that, generically, the spectrum of
quantum Hamiltonian
systems chaotic in the classical limit should conform with
the spectral properties of the random matrix ensembles.

However, the exact results presented above show that
for quantum graphs with a finite number of bonds and vertices
(finite quantum graphs) the
correspondence
with the nearest-neighbor distributions (\ref{GOENNS}) and
(\ref{GUENNS}), respectively, can
only be approximate.
Indeed, the existence of root separators implies that
the eigenvalues of the momentum, $k_{n}$, will always be
confined within the root
cells, $k_{n}\in [\hat k_{n-1},\hat k_{n}]$.
Hence, for finite quantum graphs, even though they may be
highly connected,
the statistical
distribution $P(s)$ of the nearest neighbor separations,
$s_{n}=k_{n}-k_{n-1}$,
will be restricted to the
finite domain $0<s<s_{max}$, and will not have the
characteristic long tail of the
nearest-neighbor distributions (\ref{GOENNS}) and
(\ref{GUENNS}), respectively.
This general property of the spectra of scaling quantum graphs also
follows from the
fact that their spectral function, $\Delta_{R}(k)$, is an almost
periodic function
of the momentum, and hence its zeros form an almost periodic
set \cite{HB}. It is
clear, therefore, that the distances between neighboring points
of this set are bounded,
i.e. $s<s_{max}$, and $P(s)$ is zero for $s>s_{max}$.
These observations, of course, do not preclude the possibility
that certain finite, highly connected quantum graphs are well,
or indeed even exactly described by the finite matrix
ensembles \cite{Haake}.

However, the higher the degree
of irregularity of a quantum graph,
the larger
$s_{max}$. Going upwards in the ``hierarchy of separators''
leads to an increase
in the allowed nearest neighbor spacings,
since the maximal possible distance
between neighboring separators grows by one unit of mean
spacing when going from
complexity level $m-1$ to complexity level $m$.
The mechanism for the increase of the allowed
maximal nearest-neighbors spacing as a function of $m$
is illustrated in Fig.~8.
\begin{figure}
\begin{center}
\includegraphics{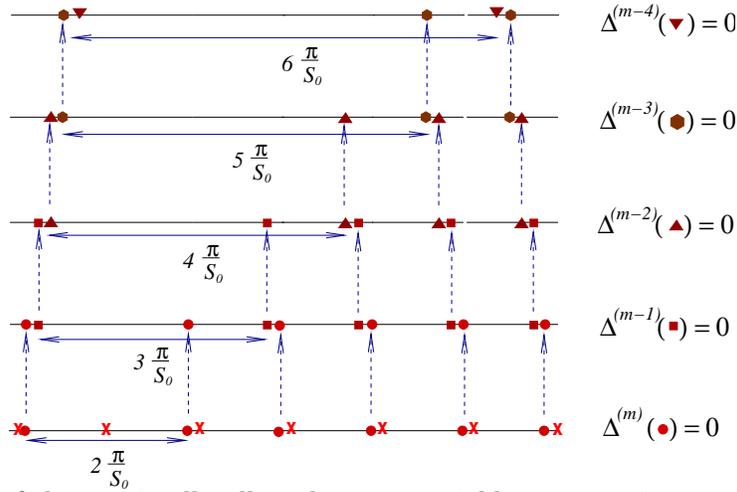}
\caption{\label{Fig.8}
Increase of the maximally allowed
nearest-neighbor separation as a function of $m$,
generated by the hierarchy of the separators.
At the regular level ($\Delta^{(m)}$),
the periodically spaced
separators $\hat k_{n}^{(m)}$ (marked
by $\times$) separate the roots
of $\Delta_{R}^{(m)}(k_{n}^{(m)})=0$ (full circles).
Their maximally allowed distance is
$2\frac{\pi}{S_{0}}$. The second order
separators (the roots of
$\Delta_{R}^{(m-1)}(k_{n}^{(m-1)})=0$), may be maximally as
far as $3\frac{\pi}{S_{0}}$ apart, etc.
The higher the hierarchy of the separator,
the larger the maximally allowed spacing of nearest
neighbors.}
\end{center}
\end{figure}
Figure~8 also shows that the roots
of a spectral equation $\Delta_{R}(k)=0$ with
irregularity degree $m$, may be no more
than $(m+1)\pi/S_{0}$ apart. This
provides a simple rule for finding an upper limit for $s_{max}$,
\begin{equation}
s_{max}^{(m)}\leq
d_{max}^{(m)}=\frac{\pi}{S_{0}}(m+1).
\label{estimate}
\end{equation}
Clearly, the possibility of having large
separations between the nearest neighbors
is necessary for producing a statistical
distribution for $s_{n}=k_{n}-k_{n-1}$ that
resembles a Wignerian distribution profile,
similar to the ones which were numerically
obtained in Ref.~\cite{QGT2}.

On the other hand, it is essential to realize that a
high irregularity degree $m$ is not enough
to guarantee Wignerian-like statistics of
the nearest neighbor spacings. A simple
numerical experiment with the
spectral equation (\ref{4vertex}) shows that
the separations between nearest neighbors do not
necessarily assume the largest possible
values (\ref{estimate}). Hence the degree of
irregularity indeed provides only an upper
limit for the nearest-neighbor separations, and
does not determine by itself their actual
values.

For example, the dressing parameters of a quantum
network can be changed continuously so
that the system undergoes a transition from an irregularity $m$ regime
to an irregularity $m+1$ regime.
As this transition happens, the roots of the spectral equation do not
respond to produce an abrupt increase of the nearest-neighbor
separations by
$\pi/S_{0}$. Instead, the maximal nearest-neighbor separation
increases smoothly as a function
of the dressing parameters.

There is a convenient way to illustrate this
increase for the four-vertex chain network, using
the structure of its spectral regime diagram (Fig.~7).
As shown in Fig.~7,
the parameter regions that
correspond to different irregularity degrees for this
graph form a system of nested diamond
shapes, with high irregularity regimes concentrating
toward the corners of the diagram. A
specific set of the action length values,
$S_{1}^{0}$, $S_{2}^{0}$, $S_{3}^{0}$, defines the
frequencies in (\ref{4vertex}) and hence the maximal
irregularity degree $m_{max}$, i.e. the total
number of diamond-shaped regions, while a choice
of the reflection coefficients, $r_{2}$
and $r_{3}$, puts the system onto a particular
point in the diagram.
Hence, one can study the effect of increasing irregularity
by traversing the spectral
regime diagram from its center ($r_{2}=r_{3}=0$) to
one of the corners (say, $r_{2}=r_{3}=1$)
along the line $r_{2}=r_{3}=r$, $0\leq r\leq 1$.
For each value of $r=r_{2}=r_{3}$ that corresponds
to a particular irregularity degree, $m$, one
can obtain numerically the maximal separation
distance, $s_{max}$, between the nearest neighbors,
and then follow its change as $m$ increases.

In addition to the maximal separation $s_{max}$
there also exists a minimal separation $s_{min}$.
The vertical bars in Fig.~9 represent the
possible range of nearest-neighbor spacings
$s_{min}\leq s\leq s_{max}$ for given $m$.
Clearly, the maximal root
separation is increasing with growing $m$. However
the increase is slower than the one given by
the linear estimate $d_{max}^{(m)}$ in Eq.~(\ref{estimate}).
\begin{figure}
\begin{center}
\includegraphics{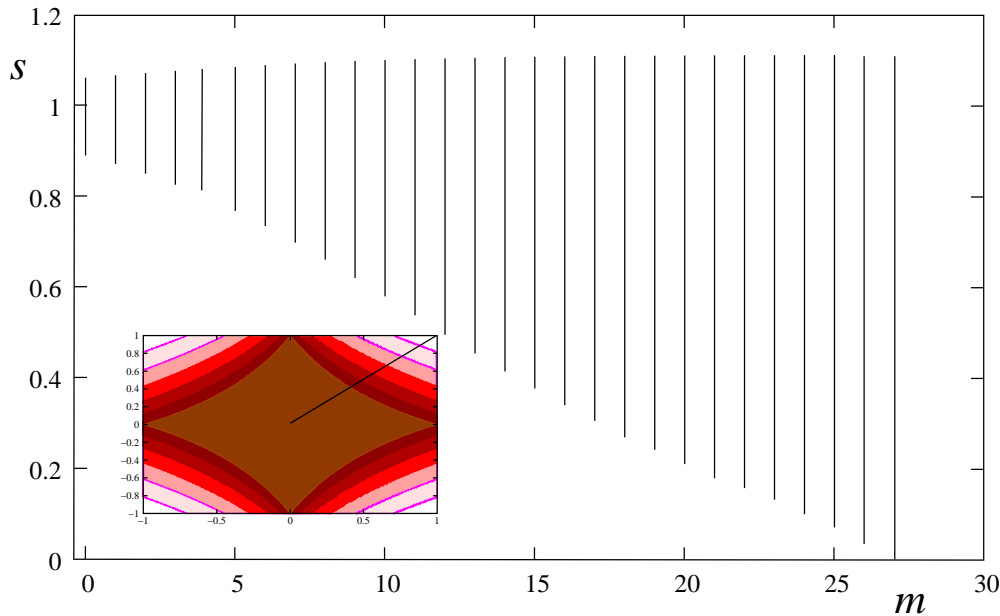}
\caption{\label{Fig.9} Range of nearest-neighbor spacings
$s/\pi$ as a function
of the irregularity degree $m$ for the
four-vertex chain. The maximal separation was obtained
based on the roots found
in the interval $0\leq k\leq \pi 10,000$ in each $m$ regime.
The bond action lengths are $S_{1}^0=0.1$,
$S_{2}^0=0.8999$, $S_{3}^0=0.0001$, which produce a
maximal irregularity degree of $m_{max}=27$.}
\end{center}
\end{figure}
Since the spectral equation (\ref{4vertex}) is an
almost periodic function, the maximal root
separation found on a sufficiently large finite interval
of the momenta (large compared to the
smallest almost-period of the function (\ref{4vertex})) is
indeed the maximal root separation
produced by this function on arbitrary intervals.

It is also important to notice that the maximal
nearest-neighbor separations $s_{max}$ can be
different for two graphs with the same
degree of irregularity. Moreover, two quantum graphs with
the same irregularity degree may have completely different
spectral statistics. This can be seen
from comparing the cases of the topologically simple four-vertex
chain graph with the fully connected four-vertex
quadrangle. The spectral statistics provided by the
latter example were previously discussed
in Ref.~\cite{QGT2}. It was shown that the nearest-neighbor
distribution follows quite closely the
anticipated Wignerian shape (both in the GOE and in the GUE cases).

The analysis of
the spectral equation of a four-vertex quadrangle with no bond
potentials and comparable bond lengths
(the case considered in Ref.~\cite{QGT2}) produces an irregularity
degree that usually does not exceed
$m\approx 25$.
This level of irregularity can be easily achieved by the
four-vertex chain, which, unlike the quadrangle,
does not produce the characteristic Wignerian
distribution profile. The distribution produced by a
four-vertex chain in different irregularity regimes is
shown in Fig.~10.
\begin{figure}
\begin{center}
\includegraphics{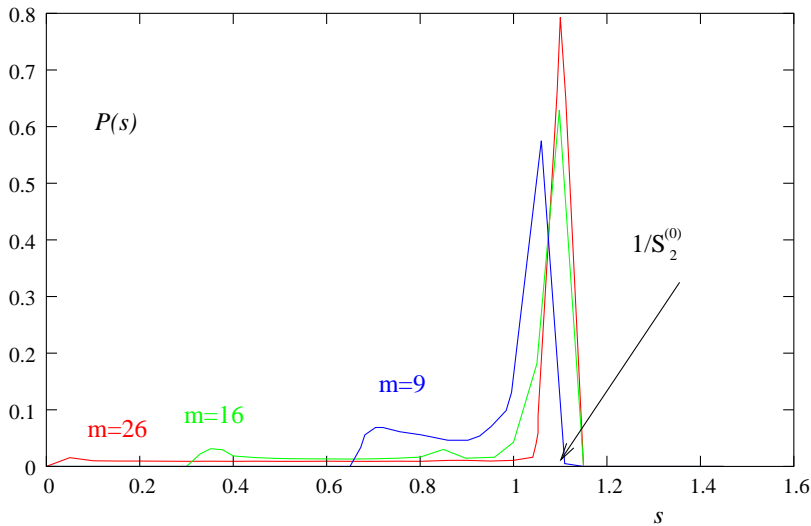}
\caption{\label{Fig.10}
Nearest-neighbor spacing probability distributions
for the eigenvalues of the four-vertex
linear-chain quantum graph with bond action lengths
$S_{1}^0=0.1$, $S_{2}^0=0.8999$, $S_{3}^0=0.0001$,
in different regimes of irregularity $m$ as a function
of spacing $s$ (in units of $\pi$).
The distribution profile is not Wignerian.
The higher the irregularity index $m$, the higher
the peak of the distribution around $s/\pi\approx
1/S_{2}^{0}$.}
\end{center}
\end{figure}
Some general features of these distribution
curves can be easily explained with the help of
elementary quantum-mechanical arguments applied to
the four-vertex chain. Indeed, it is
clear from Fig.~7, that high irregularity degree
for a four-vertex chain can be achieved
by selecting both reflection coefficients $|r_{2}|$
and $|r_{3}|$ close to $1$. Physically,
such a choice implies that the bonds of the chain are
essentially isolated, since the particle
almost never transmits from one bond to another.
This ``bond decoupling'' also manifests itself
in the spectral properties of the system by the
emergence of three apparent sub-sequences of eigenvalues,
each associated with one of
the ``isolated-bond spectra'', $\pi n/S_{i}$. Hence, one would
expect that in the case $|r_{2}|$, $|r_{3}|\approx 1$ the
nearest-neighbor separations will mostly
concentrate around the values determined by the inverse
bond lengths, $s\approx |(\pi n_i/S_i)-(\pi n_j/S_j)|$,
where $n_{i}$, $n_{j}$
are independent integers, rather than
around a peak defined by the Wignerian distribution.

Overall, the results of the statistical analysis of
the four-vertex chain spectrum show that both
the small-$s$ and the large-$s$ ends of the $P(s)$
distribution profile change slowly with
increasing irregularity degree. Even in the
case of high
irregularity degree, the behavior of the
roots of Eq.~(\ref{4vertex}) is too restricted
by the simple analytical nature of Eq.~(\ref{4vertex})
to exploit
the possibility of getting as close to, or as far
from, one another as is allowed by the hierarchy
of the separators.

Since the irregularity hierarchy presented in previous
sections is a completely general structure,
the irregularity degree $m$ produced by this scheme is a
very general index. However it does not, by itself,
determine the spectral characteristics of a given quantum graph.
While a small irregularity
degree can be provided only by a few classes of graphs
with relatively simple geometry, a large degree of
irregularity can be shared by a wide variety of graphs,
which include both the topologically simple
ones (with appropriate dressings) and the topologically
elaborate networks. It is natural, therefore,
to expect that the statistical spectral properties
produced by topologically simple graphs can differ
from the ones produced by topologically complex networks,
even if
they are characterized by the same degree of
irregularity $m$ in the sense of the bootstrapping
scheme presented above.

The four-vertex chain graph, whose spectral equation (\ref{4vertex})
contains only four oscillating
terms, is certainly too simple to produce
random-matrix-like behavior,
whereas a four-quadrangle, whose spectral
equation written in the form (\ref{4vertex}) contains about 830
terms, is already sufficiently complex.
This situation emphasizes the fact that the general
phenomenological statement ``classical chaos implies
Wignerian statistics'', implicitly
assumes sufficient complexity of the underlying classical system.

From the opposite perspective, it may be considered a
curiosity that simple networks, such as the
four-vertex chain, are capable of producing highly irregular
spectra. It is interesting in this context
to look for a more refined scheme that could distinguish
between the complexity of the spectra provided
by simple graphs (e.g. linear chains) and
the spectra of more complicated
networks, which are capable of producing
random-matrix-like spectral statistics.

In this section we studied spectral properties of quantum
graphs only as far as relevant in connection with our new
``$m$-scheme''. Much more is known about
the spectral properties of quantum graphs in general
(see, e.g.,
\cite{QGT1,QGT2,QGreview,SevTan,BerKea,Tanner1,Tanner2,Berko1,%
GnuAlt,BarGas,BSW}).
In particular the thrust in the investigation of
spectral properties nowadays is on understanding spectral
correlation functions \cite{QGT1,QGT2,QGreview,Berko1,GnuAlt,BSW} and even
deriving explicit formuls for them \cite{BerKea,BarGas}.

\section{Lagrange's inversion formula}
The periodic orbit expansions presented in Sec.~IV
are not the only way to obtain the spectrum of regular quantum graphs
explicitly. Lagrange's inversion formula \cite{LagInv}
offers an alternative route. Given an implicit equation of the form
\begin{equation}
x = a + w\varphi(x)
\label{Lagr1},
\end{equation}
Lagrange's inversion formula determines a root
$x^*$ of Eq.~(\ref{Lagr1}) according to
the explicit series expansion
\begin{equation}
x^*=a+\sum_{\nu=1}^{\infty}\, {w^{\nu}\over \nu !}\,
{d^{\nu-1}\over dx^{\nu-1}}\, \varphi^{\nu}(x)\Big|_a,
\label{Lagr2}
\end{equation}
provided $\varphi(x)$ is analytic in an open interval $I$
containing $x^*$ and
\begin{equation}
|w|\ <\ \left|{x-a\over \varphi(x)}\right|\ \ \forall\ x\in I.
\label{Lagr3}
\end{equation}
Since
the regularity condition (\ref{reg}) ensures that
the condition
(\ref{Lagr3}) is satisfied, we can use
Lagrange's inversion formula (\ref{Lagr2}) to compute
explicit solutions of regular quantum graphs.

In order to illustrate Lagrange's inversion
formula we will now apply it to the solution of
Eq.~(\ref{3hydra}).
Defining $x=S_0 k$, the $n$th root of
Eq.~(\ref{3hydra}) satisfies the implicit equation
\begin{equation}
x_n = \pi n + (-1)^n\,  \arcsin[r\sin(\rho x_n)],
\label{impl}
\end{equation}
where $\rho=S_1/S_0$ and $|\rho|<1$.
Choosing $S_0=0.3+0.7/\sqrt{2}$,
$S_1=0.3-0.7/\sqrt{2}$ and
$r=(\sqrt{2}-1)/(\sqrt{2}+1)$,
we obtain
$x_1^{({\rm exact})}= 3.26507\ldots$,
$x_{10}^{({\rm exact})}= 31.24664\ldots$ and
$x_{100}^{({\rm exact})}= 313.98697\ldots$.
We now re-compute these values using
the first two terms in the expansion (\ref{Lagr2}).
For our example they
are given by
\begin{equation}
x_n^{(2)}=\pi n + \arcsin[r\sin(\rho\pi n)]
\left\{(-1)^n + {r\rho\cos(\rho\pi n)\over
\sqrt{1-r^2\sin^2(\rho\pi n)}}\right\}.
\label{Lagexpl}
\end{equation}
We obtain $x_1^{(2)}=3.26502\ldots$,
$x_{10}^{(2)}=31.24650\ldots$ and
$x_{100}^{(2)}=313.98681\ldots$,
in very good agreement with
$x_1^{({\rm exact})}$,
$x_{10}^{({\rm exact})}$ and
$x_{100}^{({\rm exact})}$.

Although both Eq.~(\ref{regroots}) and Eq.~(\ref{Lagr2}) are
exact, and, judging from our example,
Eq.~(\ref{Lagr2}) appears to converge very quickly,
the main difference between Eq.~(\ref{regroots}) and
Eq.~(\ref{Lagr2}) is that no physical insight is gained from
Eq.~(\ref{Lagr2}), whereas Eq.~(\ref{regroots}) is tightly connected
with the classical mechanics of the graph system providing,
in the spirit of Feynman's path integrals,
an intuitively clear picture of the physical processes
in terms of a superposition of amplitudes associated
with classical periodic orbits.

\section{Discussion}
The first announcement of explicit periodic-orbit expansions
of the spectrum of regular
quantum graphs \cite{Prima} was universally
met with disbelief
and puzzlement. It seemed impossible to obtain explicit
solutions for a quantum
system that had been shown to be an excellent model of
quantum chaos \cite{MODEL}
and, moreover, is completely stochastic
in its classical limit \cite{QGT1}.
However, we found that the rejection of our results was almost
always based on the common misconception of the
``unsolvability'' of chaotic systems. We point out here
that it is not true that classically chaotic
systems are necessarily unsolvable. We hope that
this insight will eliminate
much of the reservations commonly expressed toward our results.

Examples of explicitly solvable chaotic systems are
readily available.
The shift map \cite{Devaney,Ott},
\begin{equation}
   x_{n+1}=(2x_n)\ \ {\rm mod}\ 1,\ \ x_n\in{\rm{\bf R}},
   \ \ n=0,1,2,\ldots ,
\label{shiftmap}
\end{equation}
for instance, is ``Bernoulli'' \cite{Ott},
the strongest
form of chaos. Nevertheless the shift map is readily solved
explicitly,
\begin{equation}
     x_n=(2^n\, x_0)\ \ {\rm mod}\ 1,\ \ x_n\in{\rm{\bf R}},
     \ \ n=0,1,2,\ldots .
\label{shiftsol}
\end{equation}
Another example is provided
by the logistic mapping
\begin{equation}
  x_{n+1}=\mu x_n(1-x_n),\ \ x_n\in [0,1],\ \
  0\leq \mu\leq 4,
  \ \ n=0,1,2,\ldots ,
\label{logisticmap}
\end{equation}
widely used in population dynamics
\cite{Devaney,Ott,May}. For
$\mu=4$ this mapping is equivalent with the shift map
\cite{UlNeu}
and therefore completely chaotic.
Yet an explicit
solution, valid at $\mu=4$, is given by\cite{UlNeu}:
\begin{equation}
x_n=\sin^2\left(2^n\arcsin\sqrt{x_0}\,\right),\ \ x_0\in [0,1].
\label{logisol}
\end{equation}
Therefore, as far as classical chaos is concerned, there is
no basis for the belief that classically chaotic systems do
not allow for explicit analytical solutions.
Our contribution in this paper is to show that scaling
quantum graphs provide the first examples of explicitly
solvable quantum stochastic systems.

In this paper we focussed on scaling quantum graphs
mainly because of their mathematical simplicity.
However, we will show now that for some
physical systems the scaling property arises
naturally as a consequence of the underlying physics.

Consider a taut string of length $L$, clamped
at both ends, with a piecewise constant
mass density $\mu(x)=\epsilon(x)\mu_0$,
$\epsilon(x)=\epsilon_i$, $x_{i-1}<x<a_i$,
$i=1,\ldots,4$, $a_0=0$, $a_4=L$, where
$\mu_0$ is the average mass density of the string.
This system contains the same physics as a
four-vertex linear scaling quantum graph since
the transverse acoustic excitations of the string
satisfy the same spectral equation as a
four-vertex linear scaling quantum graph. The
reason is the following.
For small transverse oscillations
the string obeys the wave equation
\begin{equation}
\left[ {d^2\over dx^2}+\omega^2{\mu(x)\over T}\right]
     \xi(x)=0,
\label{stringeq1}
\end{equation}
where $\xi(x)$ is the amplitude of the transverse
acoustic field of the string at point $x$ and $T$
is the tension in the string.

Equation (\ref{stringeq1}), supplemented with the
boundary condition $\xi(0)=\xi(L)=0$, can be
written in the form (\ref{schred}) of a
four-vertex scaling quantum graph.
Defining $E=\omega^2\mu_0/T$, we obtain
\begin{equation}
\left\{ {d^2\over dx_i^2} + \beta_i^2 E\right\}\,
 \psi_i(x_i)=0,
\label{stringeq2}
\end{equation}
where $\beta_i=\sqrt{\epsilon_i}$, $x_i\in [a_{i-1},a_i]$,
$\psi_i(x_i)\equiv \xi(x)$, $a_{i-1}<x<a_i$.
It is obvious that a web of taut strings with
more complex connectivity as in our example is
capable of simulating any scaling quantum graph.

Although the string model has not yet been realized
experimentally, a different model has been
implemented recently in the laboratory \cite{Grexp}.
This experiment models a quantum graph with the help of
interconnected microwave wave guides.
The experimental conditions are arranged such that
only the TEM mode \cite{Jackson}
can propagate in a frequency range from about
$100\,$kHz to $16\,$GHz. This allows
the authors of Ref.~\cite{Grexp} to study the
spectral properties of these microwave graphs in
great detail. Even the time-reversal violating case
is realized with the help of Faraday isolators
\cite{Mar1,Mar2}.
We suggest here that the authors of
Ref.~\cite{Grexp} could easily modify their experimental
set-up to include the case of scaling quantum graphs
in their measurements. This is done by filling the
coaxial cables representing the edges of the
quantum graphs with dielectrics of
different dielectric constants $\epsilon$, respectively.
This simple modification would allow the authors
of Ref.~\cite{Grexp} to extend the set of
experimentally accessible wave graphs enormously.
In addition to the analogues of ``conventional''
quantum graphs (quantum graphs without additional potentials
on the graph edges),
they would also be able ot study the spectral characteristics
and periodic-orbit structure of general scaling
microwave graphs, which are the analogues of scaling
quantum graphs.

The paper by Berkolaiko and Keating \cite{BerKea} is relevant in
the context of arriving at explicit formulas for physical and
mathematical characteristics of quantum graphs.
Berkolaiko and Keating's result \cite{BerKea}, however, pertains to
arriving at an explicit formula for the spectral form factor
$K(\tau)$ \cite{QGT2,BerKea}, whereas the
central result of our paper is to present
explicit formulas for the spectrum itself.
In addition the results of Berkolaiko and Keating are derived
for the special case of conventional, undressed star graphs,
whereas our formulas hold for a more general class of
dressed quantum graphs without restriction of the graph topology.
Therefore the methods and the
physical quantities computed in Ref.~\cite{BerKea} are fundamentally
different from the methods and physical quantities computed
in our paper. This also gives us the opportunity to clarify a
common confusion. It has been suggested to us that our method of
separators is the same as the method of partitions used in
the paper by Berkolaiko and Keating \cite{BerKea}, when in fact
these two methods have nothing in common. Our separators
are real numbers which isolate spectral points. The partitions
used by Berkolaiko and Keating are combinatorial entities related
to the number of ways one can represent an integer as a sum of
other integers. Partitions are a highly interesting
mathematical subject, and the greatest mathematicians, including
the famous Indian mathematician Ramanujan \cite{Rama} have
proved deep theorems about them. However, it is clear that
both methods are completely different, since even from the outset
the mathematical categories of the quantities involved are different.

Of particular importance for our investigations is the paper by
Barra and Gaspard \cite{BarGas}. These authors arrive at an
explicit formula for the nearest-neighbor spacing distribution
$P(s)$ of quantum graphs. Even more. Since the methods of Barra
and Gaspard are only based on the quasi-periodicity of the
spectral equation, their results apply to all quantum systems
with a quasi-periodic spectrum, for instance
to the dressed quantum graphs
discussed in this paper.
Since our methods yield explicit formulas for the spectral
eigenvalues themselves, we hope to be able, in future work, to
present alternative explicit representations of $P(s)$ based on
our explicit periodic-orbit expansions of the spectrum.

The standard tool of the semiclassical theory used for
studying quantum chaotic spectra is the periodic orbit
expansion for the density of states. Using the density
of states approach, the individual energy
levels are obtained
indirectly, typically with semiclassical
accuracy, as the singularities of the periodic orbit sum.
For quantum graphs, however, it
turns out that one can go one step
further, and express the
{\it individual} quantum energy levels $E_n$ in
terms of exact, explicit formulas.
Moreover, energy levels can be targeted
and labeled individually
and computed individually without the necessity
of knowing
any of the preceding energy levels.
In addition we showed that we can assign
a unique degree $m$ to any given quantum
graph, where $m$ defines the
minimum number of differentiations of
the spectral determinant necessary to reach
the regular level, which bootstraps the spectrum.
Thus quantum graphs
appear to have a certain intrinsic
degree of complexity which is
characterized by $m$.

As discussed in Ref.~\cite{Stanza}, in order to
obtain the expansion (\ref{regroots}) for
a generic quantum graph, one needs
to obtain the piercing average
of the spectral staircase, which, in general, is
a complicated task. The proposed scheme for
bootstrapping the spectrum represents a
convenient way to circumvent this
problem, and in addition it provides a
new and unexpected perspective on
the spectra of quantum graphs
by allowing
to compare their complexities.

The expansion (\ref{regroots}) is similar in spirit
to the well-known EBK semiclassical
quantization formula\cite{Gutzw,Haake,STOECK}.
Given the quantum number $n$, Eq.~(\ref{regroots}) provides an
{\it individual} expansion
of the corresponding energy eigenvalue $E_n$.
In the same spirit EBK theory provides
individual energy eigenvalues for
a given set of quantum numbers
by quantizing action integrals on tori.
Thus the two methods are similar in the sense that both
provide explicit values for the eigenenergies simply by
plugging an integer (or a set of integers)
into a known formula.
This superficial similarity
notwithstanding the underlying physics
of the two methods is completely different.
EBK relies on a simple, integrable structure of
the underlying classical dynamics based on (dynamical)
symmetries whereas our method of explicitly solving
for the spectrum of quantum graphs relies on
the construction of a network of spectral separators.

The complexity of
the expansion (\ref{regroots})
compared to the EBK quantization formula reflects the
complexity of the classical periodic orbit structure
of quantum graphs.
Moreover, the solution scheme shown above
demonstrates that the spectral
complexity of quantum graphs
can be {\it qualitatively} different for
different quantum graphs.
According to this scheme, resolving the
irregular spectra may not
amount to something as simple as
redefining the expansion coefficients
and the frequencies in Eq.~(\ref{regroots}).
Hence, further generalization
and simplification of
the individual quantum eigenvalue
quantization scheme outlined above
will most likely prove to be highly nontrivial.
Apparently, one encounters a whole
hierarchy of complexities of the
quantum spectra, even for such simple systems
as the quasi one-dimensional quantum graphs.

Concluding this section we would like to make a few comments on
the comparison between our analytical methods and standard
numerical methods for computing eigenvalues of quantum graphs.
In Ref.~\cite{PREthatis} we argued that there is an important
conceptual difference between analytical and numerical methods.
For instance analytical methods, such as ours,
provide the solution
of a whole class of objects simultaneously, whereas numerical
methods address specific solutions of specific cases, one by one.
In this sense analytical solutions are much more powerful than
numerical solutions. In addition, our explicit
analytical solutions of scaling quantum graphs are exact, whereas
the accuracy of a numerical solution is bound by the word length
of the computational device used, or,
in case of ``infinite-accuracy'' algorithms,
by the time one is willing to wait for the solution.
Even if one is content with the numerical computation of finite
accuracy, finite stretches of spectral eigenvalues, there
are at least two situations,
that require auxiliary analytical input: (i) the computation
of spectral points for very large root number $n$ and (ii)
the computation of complete spectra.
A discussion of both
cases can be found in Ref.~\cite{PREthatis}.
To this discussion we would like to add the following
recent development concerning the topic of complete spectra,
which were required for
the experimental and theoretical investigations of Ref.~\cite{Vaa}.
It was argued in Ref.~\cite{Vaa} that
even a single missing state would have invalidated the experimental
results reported in Ref.~\cite{Vaa}. Certifying completeness
of the experimental spectrum was only possible with the help of
numerical support, which itself used auxiliary analytical input
to certify the completeness of the numerical spectra.
This example illustrates clearly that
the requirement of complete spectra is not some idle academic
pursuit, but that the need for complete spectra, and thus for
analytical spectral methods, occurs in real-life situations,
including experimental physics.

\section{Summary and conclusions}
In summary, we solved the spectral problem of scaling
quantum graphs by deriving explicit, exact expressions
for each individual energy eigenvalue $E_n$ of the graph.
On the level of the spectral equation
our procedure for determining the
energy eigenvalues also defines a method for
solving analytically and
explicitly a class of transcendental equations.
This in itself is surprising and may have applications
in pure mathematics, in particular
in the theory of almost periodic functions
\cite{HB}.

The authors gratefully acknowledge financial
support by NSF grant PHY-9984075.
Work at UCSF was supported in part by the Sloan and Swartz
Foundations.


\section{Appendix A: Proof of ``one root per root cell''}
Here we provide a proof for the statement
(see Sec.~II) that one and only one root
$k_n$ of Eq.~(\ref{cos1}) is found in
the root interval $\hat k_{n-1}<k<\hat k_n$,
where $\hat k_n$ are the root separators
defined in Eq.~(\ref{sep}). In order to simplify
our task we scale and shift the argument
$k$ in Eq.~(\ref{cos1}),
\begin{equation}
k\rightarrow {1\over S_0}\ (k+\pi\gamma_0),
\end{equation}
and prove without loss of generality that
\begin{equation}
 F(x) = \cos(x)-\Phi(x)=0, \ \ \
\Phi(x)=\sum_{i=1}^N\, a_i\,
\cos(\omega_i x+\beta_i),\ \ \
|\omega_i|<1,
\label{S1}
\end{equation}
has precisely one zero $k_n$ in each interval
$I_n=(\nu_{n-1},\nu_n)$,
$n\in {\rm{\bf Z}}$,
$\nu_n=n\pi$, if the regularity condition
(\ref{reg})
is fulfilled.

We start by showing that
\begin{equation}
g(x):={\left[\sum_{i=1}^N a_i\omega_i\sin(\omega_i x+\beta_i)
\right]^2\over
1-\left[\sum_{i=1}^N a_i\cos(\omega_i x+\beta_i)\right]^2 }<1
\label{gfunc}
\end{equation}
for all $x$. The proof is straightforward. Defining
$\Theta_i=\omega_i x+\beta_i$ we have
\begin{equation}
1-\left[\sum_{i=1}^N a_i\cos(\Theta_i)\right]^2 \geq
1-\left[\sum_{i=1}^N |a_i\cos(\Theta_i)|\right]^2 \geq
1-\left[\sum_{i=1}^N |a_i|\right]^2 \geq 1-\alpha^2>0,
\label{Append1}
\end{equation}
and, with Eq.~(\ref{Append1}),
$$
g(x)\leq {\left[\sum_{i=1}^N |a_i\sin(\Theta_i)|
\right]^2\over
1-\left[\sum_{i=1}^N |a_i\cos(\Theta_i)|\right]^2 }\leq
$$
$$
1+{-1+\sum_{i=1}^N |a_i|^2+\sum_{i\neq j}|a_i a_j|\
\{|\cos(\Theta_i)\cos(\Theta_j)|+|\sin(\Theta_i)\sin(\Theta_j)|\}
\over
1-\left[\sum_{i=1}^N |a_i\cos(\Theta_i)|\right]^2 } \leq
$$
\begin{equation}
1-{1-\alpha^2\over 1-\left[\sum_{i=1}^N |a_i\cos(\Theta_i)|\right]^2}< 1.
\end{equation}

\par\noindent
We now complete the proof in six steps.
\par\noindent
(i) We observe that
$|\Phi(x)|\leq \sum_{i=1}^N |a_i|<1$ for all $x$.
\par\noindent
(ii) We use (i) to show that the end
points $\nu_n$ of $I_n$
are not roots of Eq.~(\ref{S1}):
$|F(\nu_n)|=|(-1)^n-\Phi(\nu_n)|\geq 1-|\Phi(\nu_n)|> 0$.
\par\noindent
(iii) In $I_n$ we define a new variable $\xi$ according to
\begin{equation}
x=\nu_n+\xi,\ \ \ \ 0 <\xi< \pi.\ \ \
\label{S3}
\end{equation}
Inserting Eq.~(\ref{S3}) into Eq.~(\ref{S1}) we see that
in $I_n$ the spectral function
$F(x)$ is identical with
\begin{equation}
 f_n(\xi)=(-1)^n\cos(\xi) - \varphi_n(\xi),
\label{fnxi}
\end{equation}
where
\begin{equation}
\varphi_n(\xi) = \sum_{i=1}^N a_i
\cos(\omega_i\xi+\beta_i+n\pi\omega_i).
\end{equation}
\par\noindent
(iv) Because of (i) we have
${\rm sign}\, F(\nu_n)=(-1)^n$.
We use this fact to show:
${\rm sign}\, F(\nu_n)F(\nu_{n+1})=(-1)^{2n+1}=-1$.
Since $F$ is continuous, this proves that
there is at least one root of $F$ in every $I_n$.
\par\noindent
(v) According to (iii) and Eq.~(\ref{fnxi})
the roots of $F$ in $I_n$ satisfy
$(-1)^n\cos(\xi) = \varphi_n(\xi)$, or
\begin{equation}
   \xi=h_n(\xi),
\label{S4}
\end{equation}
where
$h_n(\xi) = \arccos[(-1)^n \varphi_n(\xi)]$.
Therefore, roots of $F$
are fixed points of $h_n$.
\par\noindent
(vi) In $I_n$, because of Eq.~(\ref{gfunc}):
\begin{equation}
[h_n'(\xi)]^2 = {\left[ \sum_{i=1}^N a_i \omega_i
\sin(\omega_i\xi+\beta_i+n\pi\omega_i)\right]^2 \over
1-\left[ \sum_{i=1}^N a_i
\cos(\omega_i\xi+\beta_i+n\pi\omega_i)\right]^2 }
\  <\ 1.
\label{S5}
\end{equation}
From Eq.~(\ref{S5})
we obtain
\begin{equation}
h_n'(\xi)<1$ in $I_n.
\label{S6}
\end{equation}
Because of Eq.~(\ref{S6})
it now follows immediately that Eq.~(\ref{S4})
has only a single fixed point.
This is so since (iv) guarantees the existence
of at least one fixed point $\xi^*$ of Eq.~(\ref{S4}).
But because of
Eq.~(\ref{S6}) there cannot be any other, since
Eq.~(\ref{S6}) guarantees that $|\xi-h_n(\xi)|$ increases
monotonically to both sides of $\xi^*$.
Consequently Eq.~(\ref{S4}) has one and only one fixed point.
Since, because of (v), the fixed points of Eq.~(\ref{S4})
are the roots of Eq.~(\ref{S1}) in $I_n$, we showed that
Eq.~(\ref{S1}) has precisely one root in each
root interval $I_n$.

\section{Appendix B: Convergence of periodic orbit expansions
         for individual spectral points}
Here we show that our explicit spectral formulas converge,
and converge to the correct spectral eigenvalues.
For the zeros of (\ref{cos1})
we define the spectral staircase
\begin{equation}
N(k)=\sum_{i=1}^{\infty}\, \theta(k-k_i),
\label{stair}
\end{equation}
where $\theta(x)$
is Heavyside's $\theta$ function (\ref{Heavy}).
Based on the scattering quantization
approach
it was shown elsewhere\cite{QGT1} that
\begin{equation}
N(k)=\bar N(k) + {1\over \pi} {\rm Im}\, {\rm Tr}\,
\sum_{l=1}^{\infty}\, {1\over l}\, S^{l}(k),
\label{N1}
\end{equation}
where
\begin{equation}
\bar N(k)= {S_0 k \over \pi}\ -\ (\mu+1+\gamma_0),
\label{Nbar}
\end{equation}
and $S(k)$ is the unitary scattering matrix
of the quantum graph.
Since, according to our assumptions,
$S(k)$ is a finite,
unitary matrix,
existence and convergence of Eq.~(\ref{N1}) is guaranteed
since in the eigenangle representation Eq.~(\ref{N1})
involves nothing but the Fourier sums
$\sum_{l=1}^{\infty}\, \sin(l\sigma(k))/l$, which
according to Ref.~\cite{GR}, formula 1.4411, converge
to $[\pi-\sigma(k)]/2$ mod $2\pi$.
Therefore, $N(k)$ is well-defined for all $k$.
Since $S(k)$ can easily be
constructed for any given quantum
graph\cite{QGT1,Stanza},
Eq.~(\ref{N1}) provides
an explicit formula for the staircase function
(\ref{stair}).
This expression now
enables us to explicitly compute the zeros of
Eq.~(\ref{cos1}).

In Appendix A we proved that exactly one zero $k_n$
of Eq.~(\ref{cos1}) is located in
$I_n=(\hat k_{n-1},\hat k_n)$.
Integrating $N(k)$
from $\hat k_{n-1}$ to $\hat k_n$ and taking into
account that $N(k)$ jumps by one unit at
$k=k_n$, we obtain
\begin{equation}
\int_{\hat k_{n-1}}^{\hat k_n}\, N(k)\, dk =
N(\hat k_{n-1})[k_n-\hat k_{n-1}]+N(\hat k_n)[\hat k_n-k_n].
\label{Nint}
\end{equation}
Solving for $k_n$ and using
$N(\hat k_{n-1})=n-1$
and
$N(\hat k_n)=n$,
we obtain
\begin{equation}
k_n={\pi\over S_0}\, (2n+\mu+\gamma_0)\ -\
\int_{\hat k_{n-1}}^{\hat k_n}N(k)dk.
\label{explic}
\end{equation}
Since we know $N(k)$ explicitly, Eq.~(\ref{explic}) allows us to
compute every zero of Eq.~(\ref{cos1}) explicitly and
individually for any choice of $n$.
The representation (\ref{explic}) requires no further
proof since, as mentioned above, $N(k)$ is well-defined
everywhere, and is integrable over any finite
interval of $k$.

Another useful representation of $k_n$ is obtained by substituting
Eq.~(\ref{N1}) with Eq.~(\ref{Nbar})
into Eq.~(\ref{explic}) and using
$\bar k_n=\pi [n+\mu+1/2+\gamma_0]/S_0$:
\begin{equation}
k_n=\bar k_n
\ -\ {1\over \pi}\, {\rm Im}\, {\rm Tr}\,
\int_{\hat k_{n-1}}^{\hat k_n}\,
\sum_{l=1}^{\infty}\, {1\over l}\, S^l(k)\, dk.
\label{explic'}
\end{equation}
In the eigenangle representation of the $S$-matrix
it is trivial to show by direct calculation that
integration and summation can be interchanged in
Eq.~(\ref{explic'})
and we arrive at
\begin{equation}
k_n=\bar k_n\
\ -\ {1\over \pi}\, {\rm Im}\, {\rm Tr}\,
\sum_{l=1}^{\infty}\, {1\over l}\,
\int_{\hat k_{n-1}}^{\hat k_n} \, S^l(k)\, dk.
\label{explic''}
\end{equation}
In many cases the
integral over $S^l(k)$ can be performed explicitly, which yields
explicit representations for $k_n$.

 Finally we discuss explicit representations of $k_n$ in terms
of periodic orbits. Based on the product form
of the $S$ matrix \cite{Stanza}
the trace
of $S^l(k)$
is of the form
\begin{equation}
{\rm Tr}\, S(k)^l = \sum_{j_1\ldots j_l}
D_{j_1,j_1}U_{j_1,j_2}D_{j_2,j_2}U_{j_2,j_3}\,
\ldots\, D_{j_l,j_l}U_{j_l,j_1}=\sum_{m\in P[l]}\, A_m[l] \,
\exp\left\{iL_m^{(0)}[l] k\right\},
\label{PO1}
\end{equation}
where $P[l]$
is the index set of all possible
periodic orbits
of length $l$ of
the graph, $A_m[l]$
is the weight of orbit number $m$ of length $l$,
computable from
the matrix elements of $U$,
and $L_m^{(0)}[l]$ is the reduced action
of periodic orbit number $m$ of length $l$.
Using this result
we obtain the explicit periodic orbit
formula for the spectrum in the form
\begin{equation}
k_n=\bar k_n\ -\
{2\over \pi}\, {\rm Im}\, \sum_{l=1}^{\infty}\,
{1\over l}\, \sum_{m\in P[l]}\, A_m[l]\,
{e^{iL_m^{(0)}[l]\bar k_n}\over L_m^{(0)}[l]}\,
\sin\left[{\pi\over 2S_0}\, L_m^{(0)}[l]\right].
\label{explic'''}
\end{equation}
Since the derivation of Eq.~(\ref{explic'''}) involves only
a resummation of ${\rm Tr}\, S^l$ (which
involves only a finite number
of terms), the convergence properties of Eq.~(\ref{explic''}) are
unaffected, and Eq.~(\ref{explic'''}) converges.

Reviewing our logic that took us from Eq.~(\ref{explic}) to
Eq.~(\ref{explic'''}) it is important to stress that
Eq.~(\ref{explic'''}) converges to the correct
result for $k_n$. This is so because starting from
Eq.~(\ref{explic}), we arrive
at Eq.~(\ref{explic'''}) performing only allowed equivalence
transformations.
This is an important result. It means that even though
Eq.~(\ref{explic'''}) may only be
conditionally convergent,
it still converges to the correct result,
provided the series is summed exactly as specified in
Eq.~(\ref{explic'''}). The summation scheme specified in
Eq.~(\ref{explic'''}) means that periodic orbits have to be
summed according to their symbolic lengths
\cite{Devaney,Ott} and
not, e.g., according
to their action lengths.
If this proviso is properly taken into account,
Eq.~(\ref{explic'''}) is an explicit, convergent periodic orbit
representation for $k_n$ that converges to the exact
value of $k_n$.

It is possible to re-write Eq.~(\ref{explic'''}) into
the more familiar form of summation over prime periodic orbits
and their repetitions.
Any periodic orbit $m$ of length $l$
in Eq.~(\ref{explic'''}) consists of
an irreducible, prime periodic orbit $m_{\cal P}$ of
length $l_{\cal P}$
which is repeated $\nu$ times, such that
\begin{equation}
l=\nu l_{\cal P}.
\label{ppo1}
\end{equation}
Of course
$\nu$ may be equal to 1 if orbit number $m$ is already
a prime periodic orbit. Let us now focus on the
amplitude $A_m[l]$ in Eq.~(\ref{explic''}). If we denote by
$A_{m_{\cal P}}$ the amplitude of the prime periodic orbit, then
\begin{equation}
A_m[l]=l_{\cal P}\, A_{m_{\cal P}}^{\nu}.
\label{ppo2}
\end{equation}
This is so, because the prime periodic
orbit $m_{\cal P}$ is repeated $\nu$ times, which by itself results
in the amplitude $A_{m_{\cal P}}^{\nu}$.
The factor $l_{\cal P}$ is explained in the following way:
because of the trace
in Eq.~(\ref{explic''}), every vertex visited by the prime periodic
orbit $m_{\cal P}$
contributes an amplitude
$A_{m_{\cal P}}^{\nu}$ to the total amplitude
$A_m[l]$. Since the prime periodic orbit
is of length $l_{\cal P}$, i.e.
it visits $l_{\cal P}$ vertices, the total contribution is
$l_{\cal P}\, A_{m_{\cal P}}^{\nu}$. Finally,
if we denote by $L_{m_{\cal P}}^{(0)}$
the reduced action of
the prime periodic orbit $m_{\cal P}$, then
\begin{equation}
L_m^{(0)}[l] = \nu\, L_{m_{\cal P}}^{(0)}.
\label{ppo3}
\end{equation}
Collecting the results (\ref{ppo1}) -- (\ref{ppo3})
and inserting them into
Eq.~(\ref{explic'''}) yields
\begin{equation}
k_n=\bar k_n\ -\
{2\over \pi}\, {\rm Im}\, \sum_{m_{\cal P}}\,
{1\over L_{m_{\cal P}}^{(0)}}\,
\sum_{\nu=1}^{\infty}\, {1\over \nu^2}\,
A_{m_{\cal P}}^{\nu}\,
e^{i\nu L_{m_{\cal P}}^{(0)}\bar k_n}\,
\sin\left[{\nu\pi\over 2S_0}\,
L_{m_{\cal P}}^{(0)}\right],
\label{explic'`}
\end{equation}
where the summation is over all prime periodic orbits
$m_{{\cal P}}$
of the graph and all their repetitions $\nu$. It is important
to note here that the summation in Eq.~(\ref{explic'`}) still
has to be performed according to the symbolic lengths
$l=\nu l_{\cal P}$ of the orbits.

In conclusion we note that our
methods generalize and can be used to obtain
any differentiable function $f(k_n)$ directly
and explicitly.
Integrating over $f'(k) N(k)$
we obtain
\begin{equation}
f(k_n)=nf(\hat k_n)-(n-1)f(\hat k_{n-1})-
\int_{\hat k_{n-1}}^{\hat k_n}\, f'(k)\, N(k)\, dk.
\label{Nfint}
\end{equation}
According to the same logic that led to Eq.~(\ref{explic'''}),
we obtain
\begin{equation}
f(k_n)=nf(\hat k_n)-(n-1)f(\hat k_{n-1})-
{2\over \pi}\, {\rm Im}\, \sum_{l=1}^{\infty}\,
{1\over l}\, \sum_{m\in P[l]}\, A_m[l]\,
G_n(L_m^{(0)}[l]),
\label{explicf}
\end{equation}
where
\begin{equation}
G_n(x)\ = \
\int_{\hat k_{n-1}}^{\hat k_n}\, f'(k)\,
e^{ixk}\, dk.
\end{equation}
This amounts to a resummation
since one can also
obtain the series for
$k_n$ first, and then form $f(k_n)$.

\end{document}